# Linearly-polarized few-cycle pulses drive carrier envelope phase-sensitive coherent magnetization injection

Ofer Neufeld[*]

Technion Israel Institute of Technology, Faculty of Chemistry, Haifa 3200003, Israel.
*Corresponding author E-mail: ofern@technion.ac.il

Circularly-polarized light is well-known to induce, or flip the direction of, magnetization in solids. At its heart, this arises from time-reversal symmetry breaking by the vector potential, causing inverse-Faraday or analogous physical effects. We show here that very short few-cycle pulses can cause similar phenomena even when they are linearly-polarized and off-resonant. We analyze the new effect with *ab-initio* calculations and demonstrate that it similarly arises due to broken time-reversal symmetry and is carrier-envelope-phase (CEP) sensitive. Coherently tuning the CEP causes the induced magnetism to oscillate from zero to few percent Bohr magneton within few femtoseconds. By changing the laser angle and intensity, the magnetization sign and magnitude can be controlled. Remarkably, due to the nature of the physical mechanism that relies on spin-orbit interactions and orbital currents, the magnetization survives CEP-averaging and the effect should be accessible even in the absence of CEP stabilization. Our work opens new routes for ultrafast coherent tuning and probing of magnetism and circumvents the need for circularly-polarized driving or external magnetic fields.

Over the last two decades immense and seminal work has been carried out in the field of ultrafast magnetism [1–6]. By irradiating magnetic solids with intense linearly-polarized pulses it was shown that the magnetization can be suppressed within few femtoseconds (even as fast as single femtoseconds [7]), and the various physical mechanisms behind this and related ultrafast magneto-optical effects have been, and are still being, uncovered [8–23]. It was also shown that light can cause spin transfer [7,24–27] and spin torques [28–33], and that angular momentum can transfer between the spin and phonon sub-systems under various configurations [34–36]. Circularly-polarized pulses were measured to flip magnetization directions [37–40], and recently it was even predicted that non-magnetic materials can be turned into transient magnets on femtosecond timescales [41,42], or within few hundreds of attoseconds while the laser pulse is on due to nonlinear optical interactions [43]. In unique cases, it was also shown that resonant linearly-polarized pulses can induce a transient magnetization due to spontaneous breaking of time-reversal symmetry when substantial electronic excitation occurs [44]. Since magnetism is a ubiquitous and fundamental phenomenon that carries huge importance in devices and electronics, developing methods to optically tune it and probe it on faster timescales can potentially pave way to novel magnetic memories and switches.

Recently, the field of strong-field and attoscience in solids has been gaining a lot of attention for exploring fundamental physics in condensed matter, e.g. superconductivity [45,46], correlations [47–52], phononics [53–56], and topology [57–65]. At its heart, intense laser pulses drive highly nonlinear electron dynamics within the solid band structure, leading to a plethora of phenomena such as high harmonic generation (HHG) [66–69], Floquet dressing [70–81], and photocurrent generation [82–86]. In particular, it was shown that nonlinear multi-photon interactions driven by few-cycle pulses can generate injection currents that are carrier-envelope-phase (CEP) sensitive [82,83,85]. The photocurrents appear even in inversion-symmetric materials as long as time-reversal symmetry (TRS) is broken by the laser, which causes an asymmetric injection of charge carrier in the conduction band.

Here we propose, and numerically demonstrate with *ab-initio* calculations, a new ultrafast magnetic phenomenon – solids irradiated by intense few-cycle laser pulses can develop a coherent transient long-range magnetic order within few femtoseconds. The effect appears even when employing linearly-polarized pulses that do not carry angular momentum, because for few-cycle pulses TRS can be explicitly broken in the vector potential. Physically, we show that the short pulses drive asymmetric injection currents in conduction bands (CB), just as in photocurrent generation, but also induce transverse orbital currents if the laser polarization is chosen not along a high symmetry axis. As a result, electronic angular momentum develops within few femtoseconds and causes spin flip processes due to spin-orbit coupling (SOC). The physical mechanism is thus similar to the nonlinear inverse Faraday effect [41,87,88] but driven only for half an optical cycle, and creates an effect somewhat analogous to spin injection [89,90]. The induced magnetism is CEP-dependent, paving routes to magnetization coherent control, but importantly can also be obtained in the absence of CEP-stabilized sources. Tuning the pulse intensity and polarization angle provides control over the magnetization amplitude and sign. Magnetization measurements vs. CEP could also serve as probes of ultrafast quantum dynamics inside the material.

Let us begin by describing the details of our methodology and the physical system we explore, while technical details are delegated to the Appendix. We describe interactions between coherent short laser pulses and monolayers of bismuthumane [91,92] (BiH), and the transition metal dichalcogenide $WTe_2$ [93]. In BiH bismuth atoms are arranged in a honeycomb lattice with staggered hydrogen capping along the $p_z$ orbitals that increases the size of the gap (see Fig. 1(a)), and the lattice is inversion symmetric. In $WTe_2$ the lattice has broken inversion symmetry such that bulk photogalvanic currents can be observed even if the driving field does not break TRS, which we will show below also impacts the light-induced magnetic response. These two-dimensional quantum material systems were chosen due to their numerical simplicity, large gap, large SOC, and as two primary examples of quantum materials with/without inversion symmetry; however, as we will show below the main results and physical mechanisms are general. The monolayer is then irradiated by intense few-cycle linearly-polarized laser pulses with the following vector potential:

$$\mathbf{A}(t) = c\frac{E_0}{\omega}f(t)\sin(\omega t + \phi_{CEP})\hat{\mathbf{e}}_\theta \quad (1)$$

where $E_0$ is the electric field amplitude (defined as $E(t) = -\partial_t \mathbf{A}(t)/c$), $\omega$ is the laser field frequency, $c$ the speed of light, $f(t)$ is a temporal envelope function (see Appendix), $\phi_{CEP}$ is the CEP, and $\hat{\mathbf{e}}_\theta$ is linearly-polarized unit vector pointing to angle $\theta$ measured in the $xy$ plane above the $x$-axis (which is transverse to the covalent bonds, see Fig. 1). Note that we have employed the dipole approximation that is valid in our wavelength ranges, also indicating interactions with the magnetic part of the laser field are ignored, since those should be negligible compared to the electric dipolar response. The interaction of the vector potential and the material is described *ab-initio* using time-dependent spin density functional theory (TDSDFT) [94] within the adiabatic local spin density approximation (aLSDA), which has been shown to capture ultrafast magnetization phenomena [10,13]. We directly solve in a real-space grid representation the time-dependent Kohn-Sham (KS) equations of motion, which in atomic units take the form:

$$i\partial_t |\psi_{n,\mathbf{k}}^{KS}(t)\rangle = \left(\frac{1}{2}\left(-i\boldsymbol{\nabla} + \frac{\mathbf{A}(t)}{c}\right)^2 \sigma_0 + v_{KS}(t) + v_{so}\right)|\psi_{n,\mathbf{k}}^{KS}(t)\rangle \quad (2)$$

where $|\psi_{n,\mathbf{k}}^{KS}(t)\rangle$ is the KS-Bloch (KSB) state at $k$-point $\mathbf{k}$ and band index $n$:

$$|\psi_{n,\mathbf{k}}^{KS}(t)\rangle = \begin{bmatrix}|\varphi_{n,\mathbf{k},\uparrow}^{KS}(t)\rangle \\ |\varphi_{n,\mathbf{k},\downarrow}^{KS}(t)\rangle\end{bmatrix} \quad (3)$$

with $|\varphi_{n,\mathbf{k},\alpha}^{KS}(t)\rangle$ the spin-up/down part of the spinor with spin index $\alpha$, $\sigma_0$ a 2×2 identity matrix, and $v_{KS}(t)$ the time-dependent KS potential given by:

$$v_{KS}(\mathbf{r},t) = \int d^3r' \frac{n(\mathbf{r}',t)}{|\mathbf{r}-\mathbf{r}'|}\sigma_0 + v_{XC}[\rho(\mathbf{r},t)] + v_{ion} \quad (4)$$

where the first term in Eq. (4) is the Hartree term (describing mean-field Coulombic electronic interaction) with $n(\mathbf{r},t)=\sum_{n,\mathbf{k},\alpha} w_\mathbf{k}|\langle\mathbf{r}|\varphi_{n,\mathbf{k},\alpha}^{KS}(t)\rangle|^2$ the time-dependent electron density, $w_\mathbf{k}$ the $k$-point weights, and the sum runs over occupied bands. $v_{XC}$ is the aLSDA exchange-correlation (XC) potential, which is a functional of the spin density matrix:

$$\rho(\mathbf{r},t) = \frac{1}{2}n(\mathbf{r},t)\sigma_0 + \frac{1}{2}\mathbf{m}(\mathbf{r},t)\cdot\boldsymbol{\sigma} \quad (5)$$

, with $\boldsymbol{\sigma}$ a vector of Pauli matrices, and $\mathbf{m}(\mathbf{r},t)$ the magnetization vector. $v_{ion}$ in Eq. (4) represents the interactions of electrons with lattice ions and core electrons (through a nonlocal pseudopotential [95]), and the scalar relativistic terms as well as SOC have been separated out for notational convenience in the term $v_{so}$ in Eq. (2). The ions are assumed static, which should be a good approximation for femtosecond timescale dynamics, especially for heavy atoms. For the temporal propagation an additional complex absorbing potential (CAP) is added along the edges of the $z$-axis to avoid spurious reflection (in a method similar to that used in refs. [43,64]). From the solutions of the time-propagated KSB states we can analyze physical observables of interest, including: (1) the total charge current, $\mathbf{J}(t)$, and the unit-cell integrated spin expectation value, $\langle \mathbf{S}(t) \rangle$, the $z$-part of which is the time-dependent magnetization $M(t)$. All calculations are performed with Octopus code [96]. All technical details are delegated to the Appendix.

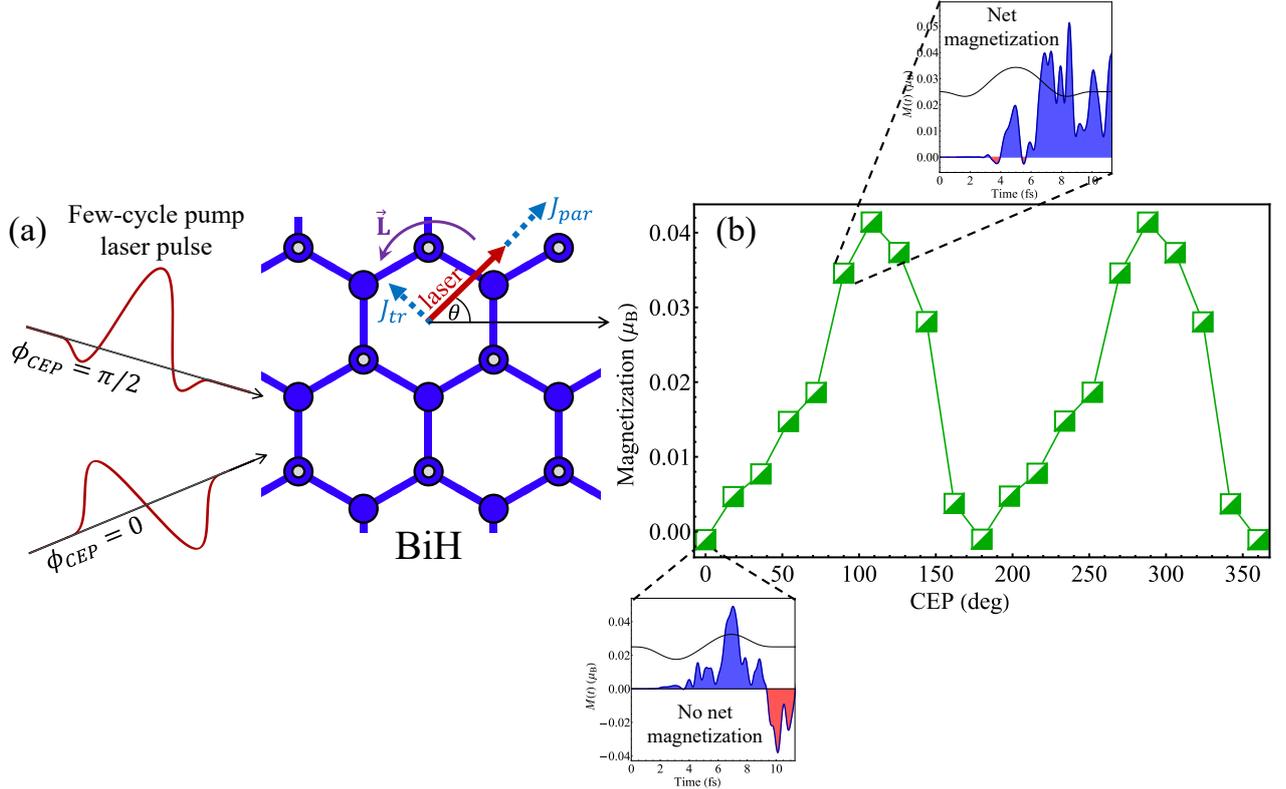

**Fig. 1.** (a) Illustration of the physical set-up and effect. An intense few-cycle laser pulse irradiates a honeycomb bismuth monolayer that is non-magnetic in its ground-state (vector potentials are illustrated in the left side of the figure). If the laser pulse breaks TRS (depending on the CEP) and is linearly-polarized along a non-high-symmetry axis (red arrow, angle $\boldsymbol{\theta}$), a net magnetization can be formed due to driving electronic angular momentum (parallel and transverse light-driven currents, see dashed blue arrows). The angular momentum induces spin flip processes enabled by SOC (purple arrow denotes electronic angular momenta which gives the sign of SOC). Due to the lattice inversion symmetry, inverting the laser direction does not change the sign of magnetization. (b) CEP-sensitive net magnetization after the laser pulse ends driven by a single cycle pulse with main carrier wavelength 3000nm, $\theta=15^0$, and intensity $4\times 10^{12}$ W/cm$^2$. Insets show the calculated temporal magnetization dynamics for two exemplary cases with $\phi_{CEP}=0$ (no net magnetization), and $\phi_{CEP}=\pi/2$ (net magnetization), with the vector potentials illustrated in black in each plot.

Let us now analyze the induced response of the spin sub-system to intense external driving with few cycle pulses. The typical set-up is illustrated in Fig. 1(a) – a laser pulse with fixed CEP excites coherent electron dynamics in the material conduction bands. The excitation and response is well known to be CEP-sensitive for such short pulses from HHG [60,97–99] and photocurrent generation set-ups [82,83,85]. Fig. 1(b) presents the main results of this letter – the light-induced magnetization in monolayer bismuth after the laser pulse ends *vs.* CEP. Magnetization dynamics for two limiting cases can be seen in insets: for $\phi_{CEP}=0$ we obtain a vanishing net magnetization, while for $\phi_{CEP}=\pi/2$ we obtain a non-vanishing strong magnetization of ~0.04 $\mu_B$. The amplitude of the net response is on par with recent calculations from circularly-polarized pulses [41,43], and is experimentally accessible with existing set-ups [7]. The magnetic response builds up over a timescale of few femtoseconds and naturally oscillates due to a broad conduction band excitation. For some CEP values the oscillation does not cancel out and leads to a non-vanishing magnetic state. We explored various laser driving parameters, and in all cases observed similar effects, indicating generality and universality (see Appendix).

In order to pin-point the physical origin of the effect, it is important to note that a net magnetic response can only be obtained as a result of spin-orbit interactions (otherwise no spin flipping occurs), and SOC can only contribute if the electronic system gains nonzero angular momentum. Thus, in order to uncover the physical mechanism there are two main questions to address: (i) How do linearly-polarized laser pulses drive electronic angular momenta if they do not carry spin angular momentum? (ii) Why do spin-flip processes build up over time instead of canceling out?

For part (i), the answer lies in the lattice geometry and the choice of driving lattice angle. Fig. 2(a) (yellow curve) shows the calculated transverse injection current after the laser pulse ends *vs.* the driving angle for the $\phi_{CEP} = \pi/2$ case driving a net magnetization. As shown, when driving along high symmetry lattice axes there is vanishing transverse current response that arises from inherent mirror-symmetry-based selection rules [100]. However, along non-high-symmetry axes non-zero transverse responses are allowed from crystal symmetry. The short temporal envelope of the pulse insures the injection of carriers is asymmetric in the Brillouin zone (BZ), meaning net Hall photocurrents are generated (as has been measured e.g. through HHG [101,102]). Together with the trivially expected parallel current responses (in Fig. 2(a), red curve), an overall electronic orbital angular momentum (OAM) is generated even if the pulses carry zero spin angular momentum. In the Appendix Fig. 5 we show that indeed short cycle pulses generate non-vanishing electronic OAM, which correlates well with the induced magnetic response. The net magnetization is also in correspondence with the multiplication of the parallel and transverse photocurrents, clearly indicating the need for their simultaneous excitation (to generate electronic OAM). Such a response is expected to be even stronger in materials with broken inversion symmetry and nonzero Berry curvatures, as we will show below for $WTe_2$.

The answer to part (ii), and physical origin of the effect, is the TRS breaking induced by the few-cycle pulse. Essentially, the short pulse can break TRS (if $\phi_{CEP}$ is chosen different than 0 or $\pi$). This is clear when looking at the laser pulse in Fig. 1(a) – the vector potential can be chosen to respect TRS, or break it, such that there are no equivalent half cycles with opposite signs driving carrier dynamics (see illustration in Fig. 1(a)). Explicitly, in such a situation (with $\phi_{CEP} = \pi/2$ and $\int \mathbf{A}(t)dt \neq 0$) the electric laser field only respects a dynamical symmetry that couples TRS with mirror axes, but breaks TRS [85,100]. Overall, net magnetization is then formally allowed [103]. In practice, the magnetic response is driven by the electronic OAM, which is generated only for one half cycle and therefore cannot cancel out by interactions with a subsequent field half cycle that does not exist, or is much weaker in power. For standard long multi-cycle pulse, TRS would be respected and dynamics driven over the multiple half cycles would cause an alternating OAM and an oscillating spin response (because with TRS one has that $\int \mathbf{A}(t)dt = 0$), rather than inducing a non-vanishing net spin flip (as established in ref. [43]).

Interestingly, while injection currents change sign upon changing the sign of the drive (by taking $\phi_{CEP} \rightarrow \phi_{CEP} + \pi$, or by taking $\theta \rightarrow -\theta$), this is not the case for the magnetic response. Indeed, upon such an inversion operation both the parallel and transverse components of the injection current changes sign, such that the electronic OAM sign is preserved (see illustration in Fig. 1(a)). Fig. 1(b) clearly shows that the induced OAM and magnetic response is symmetric under $\pi$ translation of the CEP. Figure 2(b) further shows that by changing the driving angle one can tune the sign and magnitude of the magnetic state, as that indeed flips the orbital current directionality. One positive consequence of this is that upon CEP averaging (e.g. in Fig. 1(b)) the total magnetic response survives, suggesting a CEP-stable laser source is not necessary (though that would provide enhanced tunability). Fig. 2(c) shows the induced magnetic response *vs.* the driving field amplitude in BiH, demonstrating that by increasing the laser-matter interaction strength stronger magnetism can be induced. This is in correspondence with the physical mechanism of the interaction – stronger laser driving yields a larger current response, causing larger electronic angular momenta and probability of spin flip processes.

Overall, we can connect the light-induced magnetization in a non-magnetic system to a product of the electronic OAM and the SOC strength: $M_z \propto L_z \alpha_{SO}$ (with $\alpha_{SO}$ the SOC strength). Since electronic OAM is proportional (at least heuristically) to the product of the parallel and transverse photocurrents, we expect that: $M_z \propto J_\perp J_\parallel \alpha_{SO}$ (with $J_\perp$ and $J_\parallel$ the parallel and transverse photocurrent amplitudes, respectively). In the Appendix Fig. 6 we show that the intensity scaling of the induced magnetization (as shown in Fig. 2(c)) corresponds to a scaling of $M_z \propto (E_0)^8$, i.e. to a total of eight driving photons translate to the net magnetization. In our conditions at least four photons are required to induce a photocurrent response (since a single photon has energy of ~0.41 eV, while the gap is ~1.3 eV, and an odd number of photons cannot excite shift currents), meaning the physical mechanism behind the light-driven magnetism is not proportional to just the parallel or transverse shift current excitation, but to their product (connecting to the OAM). In terms of KSB states, we could express the induced magnetization as:

$$M_z = -i \sum_{\mathbf{k}} w_{\mathbf{k}} \int dt \, \langle \psi_{n,\mathbf{k}}^{KS}(t) | [L_z, v_{so}] | \psi_{n,\mathbf{k}}^{KS}(t) \rangle \qquad (6)$$

where Eq. (6) describes the $k$-summed (with $w_{\mathbf{k}}$ the $k$-point weights) and time-integrated electronic spin-orbit torque, which is summed over the unit cell and occupied states. Here $L_z$ is the OAM operator transverse to the monolayer, which is tricky to define in periodic systems, but can for instance be defined locally at the atomic centers [104]. The spin-orbit interaction term takes the form:

$$v_{so} = \frac{-i}{4c^2} (\nabla v_{ks}(\mathbf{r}) \times \nabla) \cdot \boldsymbol{\sigma} \qquad (7)$$

, where Eq. (7) is in practice proportional to $\alpha_{SO}$. In Eq. (6) the induced magnetization then requires absorption of photons to generate nonzero $L_z$, in turn allowing spin-orbit torques and light-driven magnetism. If the pulse duration is long, the time-integral in Eq. (6) will cause the net magnetization to vanish due to oscillations in $L_z$. Even in short pulses, if $L_z$ from the different half-cycles cancels out, $M_z$ vanishes as well. This corresponds well with our *ab-initio* simulations, with the key point in this work being that $L_z$ can be generated even by few-cycle linearly-polarized pulses if the laser parameters are properly chosen. Thus, the physical mechanism behind the effect is can be imagined as an injection of spin-polarized charge carriers through the bands, which is facilitated by the cascaded excitation of electronic OAM and spin-orbit interactions.

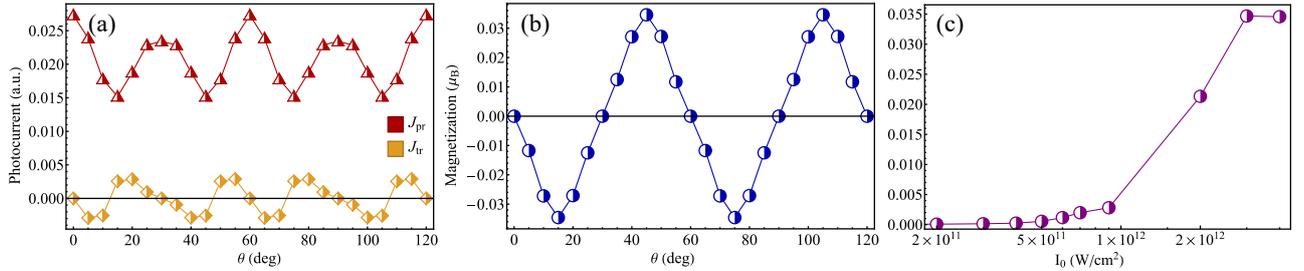

**Fig. 2.** (a) Light-driven injection currents calculated in similar conditions to Fig. 1 (but with $\phi_{CEP}=\pi/2$), showing both parallel and transverse (to the driving laser) induced responses. Both parallel and transverse currents change sign upon inverting the driving field, such that the induced electronic angular momentum and magnetization sign is maintained. (b) Light-induced magnetization *vs.* driving angle. (c) Light-induced magnetization *vs.* driving laser peak power, showing that the amplitude of the magnetization can be controlled with the laser power.

The generality of this physical mechanism implies it can be implemented in any system, e.g. initially magnetic or not, solid-state or other, inversion symmetric or not, etc. The main attributes required are a sufficiently short linearly-polarized laser pulse that breaks TRS, driving along non-high-symmetry axes, and a material with sufficiently large SOC term to induce substantial spin-flip torques.

Lastly, we explore a similar response in WTe$_2$ that does not exhibit inversion symmetry, and consequently, supports photogalvanic light-driven currents even if the driving pulse does not break TRS. Figure 3 presents similar physical phenomena with slightly weaker induced magnetic states (due to a weaker SOC in WTe$_2$ compared to bismuth), proving the generality of the effect. Notably, Fig. 3(b) shows that the magnetization can be induced by any generic value of CEP in this system – because persistent photocurrents (both parallel and transverse) can be generated here without the need to break TRS, the orbital moments are also persistent and cause non-vanishing spin flipping. Intuitively, this can be understood as a TRS breaking in the electronic dynamics themselves (since each half cycle drives a different form of orbital currents due to the broken inversion symmetry), rather than laser-induced TRS breaking. The specific CEP-dependent current also shows highly nonlinear behavior that could be indicative of the underlying ultrafast electron dynamics in the material, which might be useful for developing novel ultrafast spectroscopies of electronics and spintronics.

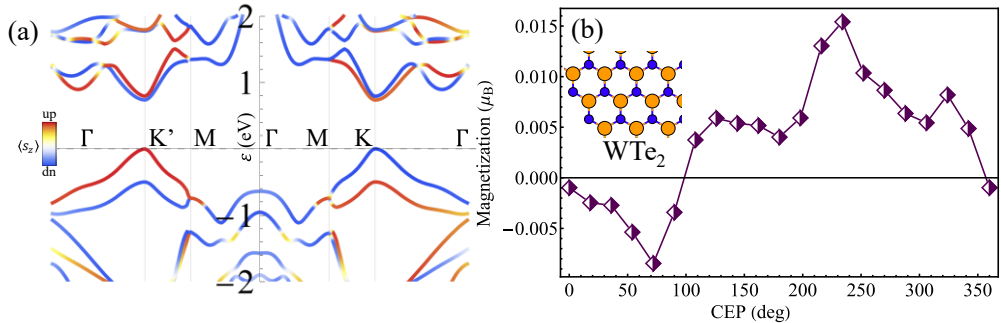

**Fig. 3.** CEP-sensitive Light induced magnetization in WTe$_2$. (a) Band structure in WTe$_2$ along high symmetry lines showing spin split bands as a result of strong SOC and inversion symmetry breaking. (b) CEP-dependent light-induced magnetization after the laser pulse driven in similar conditions to Fig. 1, but with 2000nm wavelengths, and showing that induced magnetic responses can arise for all values of CEP since WTe$_2$ does not respect inversion symmetry. Inset in (b) shows the lattice structure.

To summarize, we proposed and numerically demonstrated a new magneto-optical effect – few cycle laser pulses that are linearly-polarized (and do not carry angular momentum) can drive ultrafast magnetization (or magnetization flipping) in solids, even in systems exhibiting inversion symmetry, and which are non-magnetic in their ground state. We uncovered the physical origin of this effect by analyzing *ab-initio* calculations, and found that it originates in spin flipping phenomena that arise on a half-optical-cycle level (which are not cancelled out by subsequent half cycles), made possible by transverse current responses and light-driven electronic OAM. The effect is CEP-sensitive and can be coherently controlled with the various laser parameters to tune the sign and magnitude of the magnetization, though we showed CEP stabilization is not a necessary requirement. Our results pave the way to sub-cycle lightwave coherent control of magnetism, potentially even on attosecond timescales, which avoid the need to use circularly-polarized light or magnetic fields [105–107]. Looking forward, we expect similar control schemes to be developed using multi-chromatic pulses that permit similar TRS breaking [72,73,75,77,81,86,108–110], and for spin harmonics [111]. This effect should also be applicable for spectroscopically probing ultrafast magnetism and chirality as an alternative to circular dichroic schemes.

*Acknowlegenemts.* ON gratefully acknowledges the scientific support of Prof. Dr. Angel Rubio.

**APPENDIX**

This appendix details the numerical procedures employed in the main text, and some additional results with various laser parameters that support our conclusions. The KS equations were solved on a real space grid with spacing 0.39 Bohr for BiH and 0.38 bohr for WTe$_2$ with the shape of the hexagonal unit cell (in the *xy* plane), which was taken at the honeycomb symmetric geometry for Bih=H in ref. [91] (*a*=*b*=5.53Å, and a Bi-H distance of 1.82Å), and the experimental configuration in WTe$_2$. The *z*-axis was treated with finite boundary conditions with an added vacuum region of length 50 Bohr. We used a Γ-centered *k*-grid of 24×24 *k*-points for BiH and 36×36 *k*-points for WTe$_2$ without any assumed symmetries, which converged the nonlinear response. The equations were propagated with a time-step of 4.8 attoseconds, from which we calculated the total current and magnetic response with a method identical to ref. [43], and where the current and magnetic response where averaged over several femtoseconds after the laser pulse ends. Absorbing boundaries were employed through a CAP along the *z*-axis with a width of 15 Bohr and a maximal magnitude of 1 a.u with a sinusoidal shape [112]. The temporal field amplitude was chosen as a super-sine form for numerical convenience [113]:

$$f(t) = \left(sin\left(\pi \frac{t}{T_p}\right)\right)^{\left(\frac{\left|\pi\left(\frac{t}{T_p}-\frac{1}{2}\right)\right|}{w}\right)} \tag{8}$$

with $w$=0.75, $T_p$ the duration of the laser pulse which was taken to be very short in multiples of laser half cycles (as described in the main text).

We also present supporting results for the generality of the conclusions discussed in the main text. For instance, Fig. 4(a) shows CEP-sensitive magnetic response calculated in BiH at main carrier wavelength of 2000nm and half-cycle pulses (unlike the main text data which was calculated at 3000nm central wavelengths with single-cycle pulses). Figure 4(b) presents CEP-sensitive magnetization data for 3000nm driving pulses that are 2-cycle long (still short enough to induce an effect). In all conditions examined, we noticed a similar physical effect arising, though its magnitude and sign varied depending on conditions.

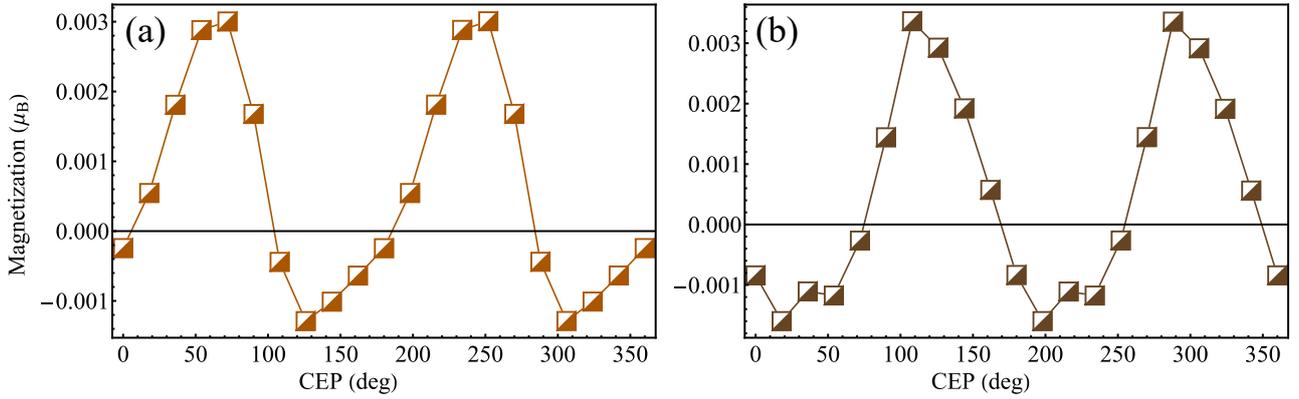

**Fig. 4.** CEP-sensitive magnetization in BiH driven by pulses in similar setting to Fig. 1(b), but with: (a) 2000nm central carrier wavelength of half a cycle duration and intensity $10^{12}$ W/cm$^2$. (b) 3000nm central carrier wavelength of two-cycle duration.

Figure 5 presents the light-induced net magnetization in the same conditions as in Fig. 1 in the main text, compared with the cell-averaged value of the OAM (an approximation for $L_z$), as well as the product of the parallel and transverse induced photocurrents. In both cases largely good agreement is obtained, supporting the scaling laws and physical mechanism discussed in the main text.

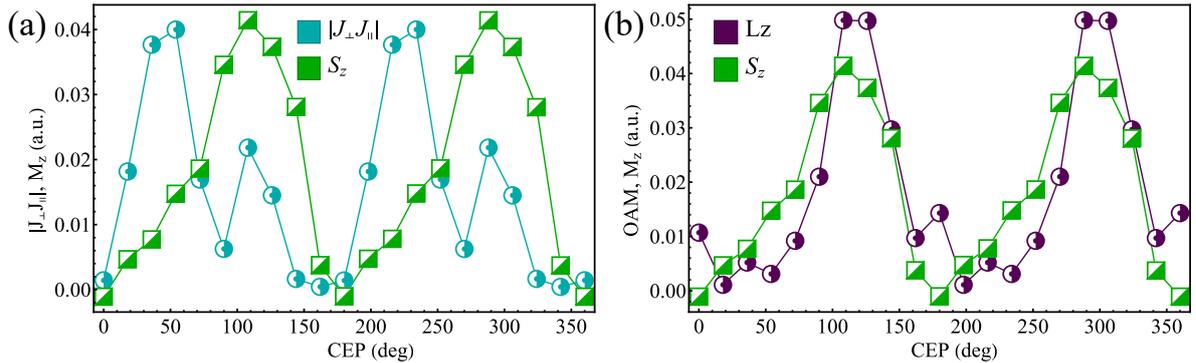

**Fig. 5.** CEP-sensitive magnetization in BiH driven by pulses in similar setting to Fig. 1(b), but compared to: (a) the product of the parallel and transverse photocurrent amplitudes, (b) the cell-averaged electronic OAM.

Lastly, Fig. 6 presents the scaling of the induced magnetization with the driving intensity as in Fig. 2(c), but with two optional scaling laws connecting to: excitation of a photocurrent in our conditions (minimum 4 photon absorption), or the product of two photocurrent excitations (parallel and transverse, both needed for nonzero OAM, requiring minimal 8 photon absorption). As seen, the scaling with $(E_0)^8$ fits much better to the numerical data, supporting our proposed physical mechanism. This prediction could be employed in experiments to test the physical mechanism behind the effect and separate it from other phenomena.

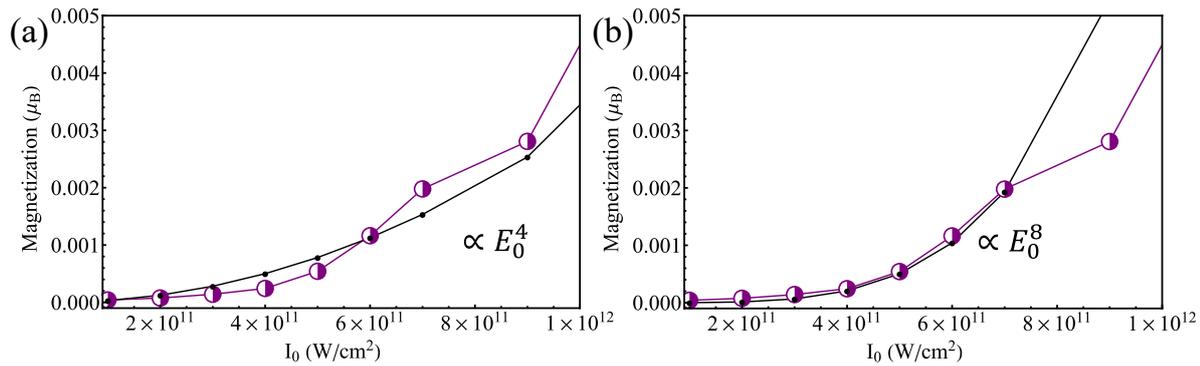

**Fig. 6.** Intensity dependence of net magnetization in BiH driven by pulses in similar setting to Fig. 2(c), but compared with two optional scaling laws: (a) optimal fit to $E^4$, (b) optimal fit to $E^8$.